\documentclass[preprint,1p,12pt]{elsarticle}

\biboptions{authoryear}

\usepackage{epsfig}
\usepackage{float}

\usepackage{amssymb}
\usepackage{amsmath}

\newcommand{\aap}{    {\it Astron. Astrophys.}}

\newcommand{\apj}{    {\it The Astrophys. J.}}

\newcommand{\solphys}{{\it Solar Phys.}}

\usepackage[colorlinks = true, linkcolor = blue, urlcolor  = blue, citecolor = blue, anchorcolor = blue]{hyperref}
\usepackage{lineno}

\journal{Advances in Space Research}
\begin{document}
\begin{frontmatter}
\title{Detailed analysis of dynamic evolution of three Active Regions before flare and CME occurrence at the photospheric level }
\author{Yudong Ye$^{1,2}$} 
\author{M. B. Kors\'os$^{3,4}$}
\author{R. Erd\'elyi$^{3,5}$}
\address{1. SIGMA Weather Group, State Key Laboratory of Space Weather, National Space Science Center, Chinese Academy of Sciences, Beijing 100190, China}
\address{2.University of Chinese Academy of Sciences, Beijing 100049, China}
\address{3. Solar Physics \& Space Plasma Research Center (SP2RC), University of Sheffield, Hounsfield Road, S3 7RH, UK}
\address{4. Debrecen Heliophysical Observatory (DHO), Konkoly Astronomical Institute, Research Centre for Astronomy and Earth Sciences, Hungarian Academy of Sciences, Debrecen, P.O.Box 30, H-4010, Hungary}\address{5.Department of Astronomy, E\"otv\"os Lor\'and University, Budapest, Hungary}

\begin{abstract}
We present a combined analysis of the applications of the weighted horizontal magnetic gradient (denoted as $WG_{M}$ in  {\color{blue}Kors\'os {\it et al.}, ApJ, 802, L21, 2015}) method and the magnetic helicity tool ({ \color{blue}Berger \& Field, JFM, 147, 133, 1984)} employed for three active regions (ARs), namely NOAA AR11261, AR11283 and AR11429. We analyse the time series of photospheric data from the Solar Dynamics Observatory taken between August 2011 and March 2012 during which period these AR have hosted 8 M- and 6 X-class flares. All three active regions produced series flares and CMEs. AR 11261 had four M-class flares where one was accompanied with a fast CME. AR 11283 had similar activity with its two M- and two X- class flares occurred, however,  only with a slow CME. Finally, AR 11429 was the most powerful  of the three active regions as it hosted five very compact solar eruptions. For applying the $WG_{M}$ method we employed the Debrecen Sunspotdata Catalog and for estimating the magnetic helicity at the photospheric level we used the Space-weather HMI Active Region Patches (SHARP's) magnetograms from SDO/HMI (Solar Dynamic Observatory/Helioseismic and Magnetic Imager). 
We followed the evolution of the components of the $WG_{M}$ and the magnetic helicity before the flare and CME occurrences. We found an unique and mutually shared behavior, called the U-shaped pattern, of the weighted distance component of  $WG_{M}$ and of the shearing component of the helicity flux before the flare and CME eruptions. This common pattern is associated with the decreasing-receding phase yet reported only known to be a necessary feature prior to solar flare eruption(s), but found now at the same time in the evolution of the shearing helicity parameter. This result leads to the conclusion that (i) the shearing motion of photospheric magnetic field may be a key driver for the solar eruption in addition to the flux emerging process, and that (ii) the found decreasing-approaching pattern in the evolution of shearing helicity may be another precursor indicator for improving the forecasting of solar eruptions.
\end{abstract}


\end{frontmatter}

\section{Introduction}

The magnetic field topology of a solar active region (AR) plays an important role in the flare and coronal mass ejection (CME) processes. Shearing motion and flux emerging are accounted for and viewed widely as responsible for such eruptive changes in the magnetic field topology of an AR. Further,  the large CMEs are often associated with more energetic flares \citep{Yashiro2006,Hudson2010} indicative of a common underlying physical mechanism of flares and CMEs.

Adjacent opposite magnetic polarities associated with the sites of large-scale eruptive events have their own strongly-sheared localised polarity inversion line (PIL) where the magnetic field gradient is high, and, which indicates the existence of the intense electric currents and therefore large free magnetic energy in the solar atmosphere \citep{schrijver2007}. The free energy often becomes the energy source of flares and CMEs. Therefore regions around PILs are preferred area of interest to search for reliable precursors of these dynamic events. A recent comparative review about the various forecasting methods and their capabilities for predicting solar eruptions can be found in \citep[see, e.g.,][and references therein]{Benz2017, Leka2017}.

The magnetic field is strongly sheared in flaring locations \citep{Hagyard1990} and large-scale is shear built up through the slow motion of footpoints stretching the length of loops \citep{Roudier2008}. In the literature, the important condition of flux emergence is more widely accepted than the shearing motion of the footpoint of AR to trigger solar flares \citep{Chandra2009,Takafumi2014,Louis2015}. A  general concept is that new emerging magnetic flux (tube) may interact with the pre-existing flux (tubes) where reconnection may occur in the current sheet, which forms between the old and new fluxes. In this process the importance of the emergence of the flux may seem to outweigh the associated shearing of the magnetic field, leading to focus by many on studying the various measures of flux emergence. By analysing magnetic helicity, especially its shearing component, we argue that shearing may provide an important clue prior to flare and CME eruption.

Magnetic helicity in a volume $V$ is defined by $H= \int_{V} {\bf A} \cdot {\bf B} dv$, where ${\bf B}$ is the magnetic field, and ${\bf A}$ is the corresponding vector potential which satisfies ${\bf B}=\nabla \times {\bf A}$. Magnetic helicity in an open volume condition like in ARs was first introduced by \cite{Berger1984b} as a description of how the magnetic field is sheared or twisted compared to a reference potential field \citep{Berger1984}. Analysing magnetic helicity provides insight into understanding the underlying mechanism of solar magnetic activities such as flare onset and of CMEs.

As a measure of the non-potentiality of the solar magnetic field, magnetic helicity can either be generated by photospheric shearing motion or be transported across the photosphere through emerging of twisted magnetic structures \citep{Zhang2012}. During the evolution of magnetic field, the total magnetic helicity conservation cannot relax to a potential field. Therefore, the accumulated magnetic helicity could be a source of a CME occurrence in a non-equilibrium state \citep{Demoulin2007, Demoulin2009}. The amount of helicity stored in pre-flare structures determines whether a big flare will be eruptive or confined \citep{Nindos2014}.

 In this article, we investigate three different ARs with the methods of the weighted horizontal magnetic gradient (denoted as $WG_{M}$) developed in \cite{Korsos2015} and a magnetic helicity analysis \citep{Berger1984,Berger1984b} for improving our flare/CME capability prediction. The application of the two methods focuses on the evolution of an active region including analysis of sunspot movements and changes in magnetic properties to improve  the potentials to predict flares and CMEs using pre-eruption parameters. All three investigated active regions, namely ARs 11261, 11283 and 11429, produced a series flares and CMEs (see for the details Tables \ref{table1}-\ref{table3}). From the three studied ARs the AR 11429 was the most powerful as it hosted five very compact solar eruptions.  In Section \ref{analyses}, we describe the detailed analysis of three ARs by applying the $WG_{M}$ method and by evaluating the evolution of their magnetic helicity, respectively. Then, Section \ref{conclusion} concludes about the dedicated complementary use of the $WG_{M}$ method and the magnetic helicity calculation in terms of the flare and CME forecasting capabilities.

\section{Analysis} \label{analyses}

\subsection{ Application of weighted horizontal magnetic gradient in 3 different ARs}

First, we investigate the pre-flare and CME dynamics of AR 11261, 11283 and 11429 with the weighted horizontal magnetic gradient (denoted as $WG_{M}$) between two opposite magnetic polarity sunspot groups introduced by \cite{Korsos2015}. The method is based on tracking changes of the solar surface magnetic configuration in ARs, as flare pre-cursors, with about an hourly resolution, with the purpose of predicting energetic flares, above M5.  In \cite{Korsos2015}, two diagnostic tools were introduced to probe the pre-flare behavior patterns. The first one is based on the relationship between the values of the maxima of the $WG_{M}$ and the intensity of the flare(s). The viability of the relationship in terms of flare forecast capability was tested on the largest available statistical sample of 61 cases observed during the SOHO/MDI era. It was concluded that this connection may provide useful insights into the relationship between the accumulated free energy, represented by $WG_{M}$  as a proxy measure, and the released energy represented by the highest GOES-class in a set of homologous flares as another proxy measure.  The second tool developed, the prediction of the flare onset time, is based on the relationship found between the duration of diverging motion of the barycenters of opposite polarities until the flare onset and duration of the compressing motion of the area-weighted barycenters of opposite polarities. These new proxies greatly enhance the capability of forecast, including {\it(i)} the expected highest intensity flare-class; {\it(ii)} the accuracy of onset time prediction and {\it(iii)} whether a flare, stronger than M5 in terms of the GOES classification scheme, is followed by another same energetic flare event(s). 

\begin{figure*}
\centering
\includegraphics[scale=0.65]{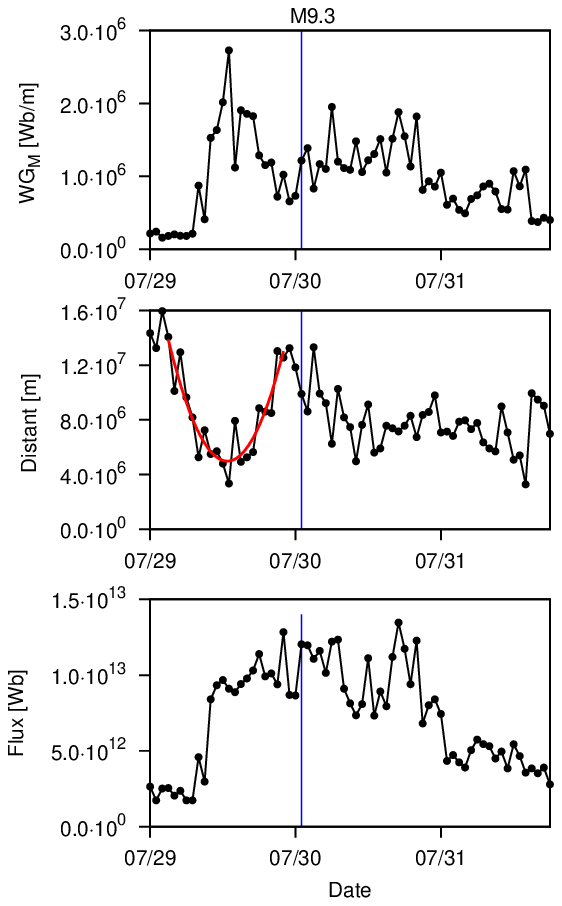}
\put(-123,190){(a)}
\includegraphics[scale=0.65]{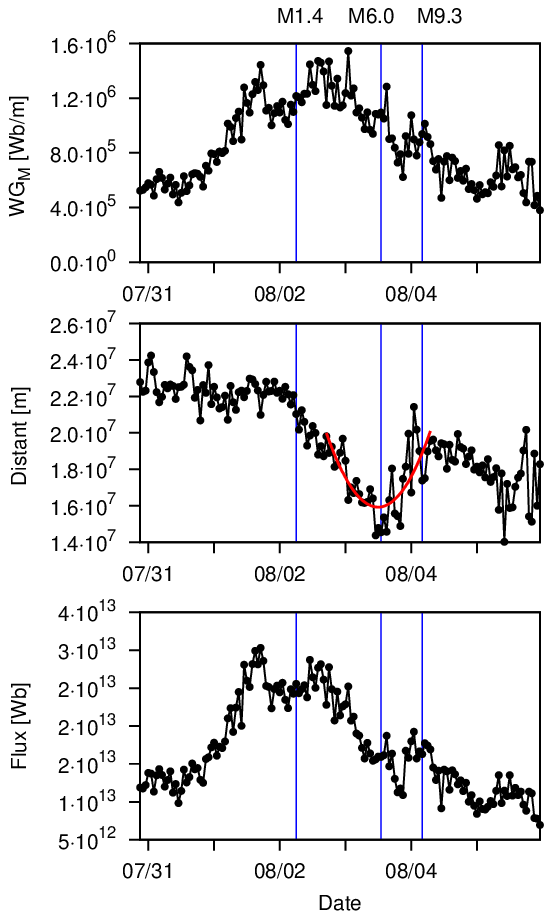}
\put(-120,190){(b)}
\includegraphics[scale=0.65]{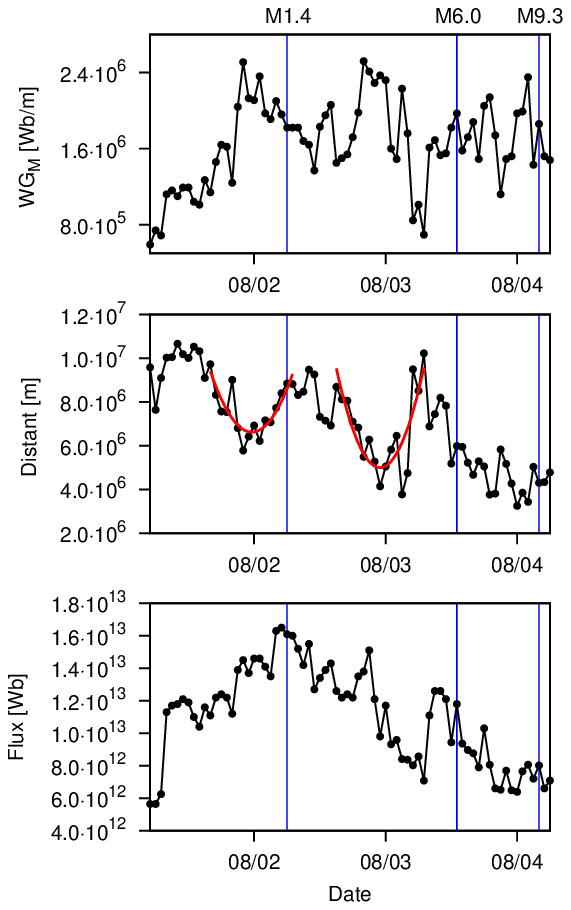}
\put(-118,190){(c)}
\includegraphics[scale=0.19]{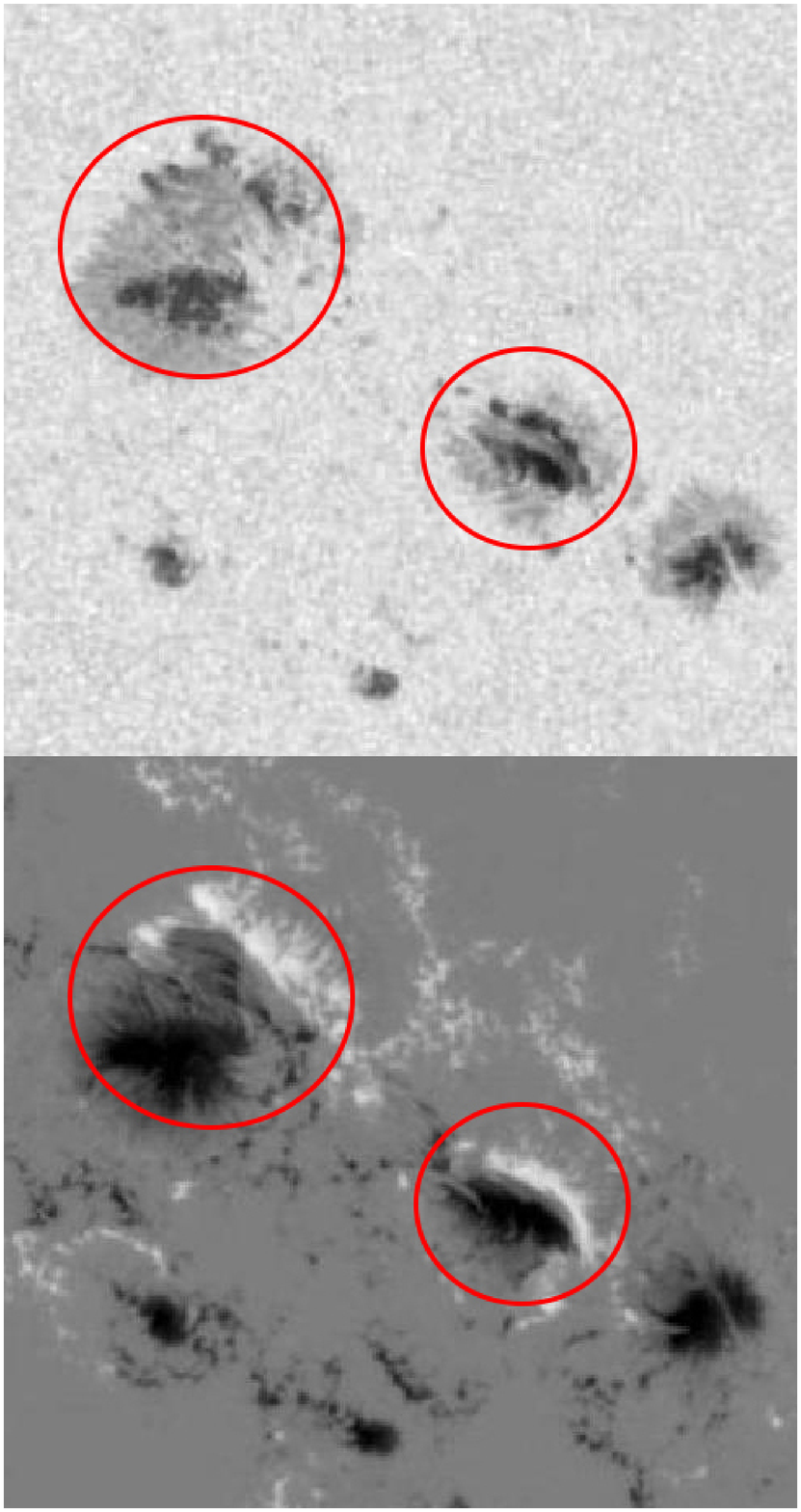}
\put(-95,190){(d)}
\put(-85,126){{\it Area 1}}
\put(-50,106){{\it Area 2}}
\put(-85,36){{\it Area 1}}
\put(-50,16){{\it Area 2}}
\put(-85,36){{\it Area 1}}
\put(-50,16){{\it Area 2}}
\caption{\label{fig11261} (a) (b) and (c): Top panel: variation of $WG_{M}$ as a function of time; Middle panel: evolution of distance between the area-weighted barycenters of the spots of opposite polarities; Bottom panel: unsigned flux of all spots in the encircled area as a function of time. (d): Top panel is intensity and bottom panel is magnetogram of AR 11261.}
\end{figure*}

\begin{figure*}
\centering
\includegraphics[scale=0.81]{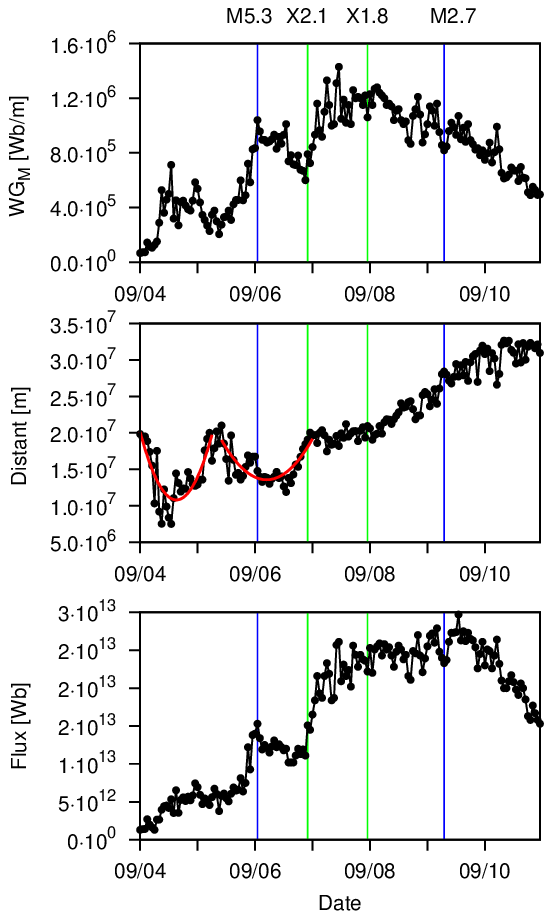}
\put(-135,215){(a)}
\includegraphics[scale=0.3]{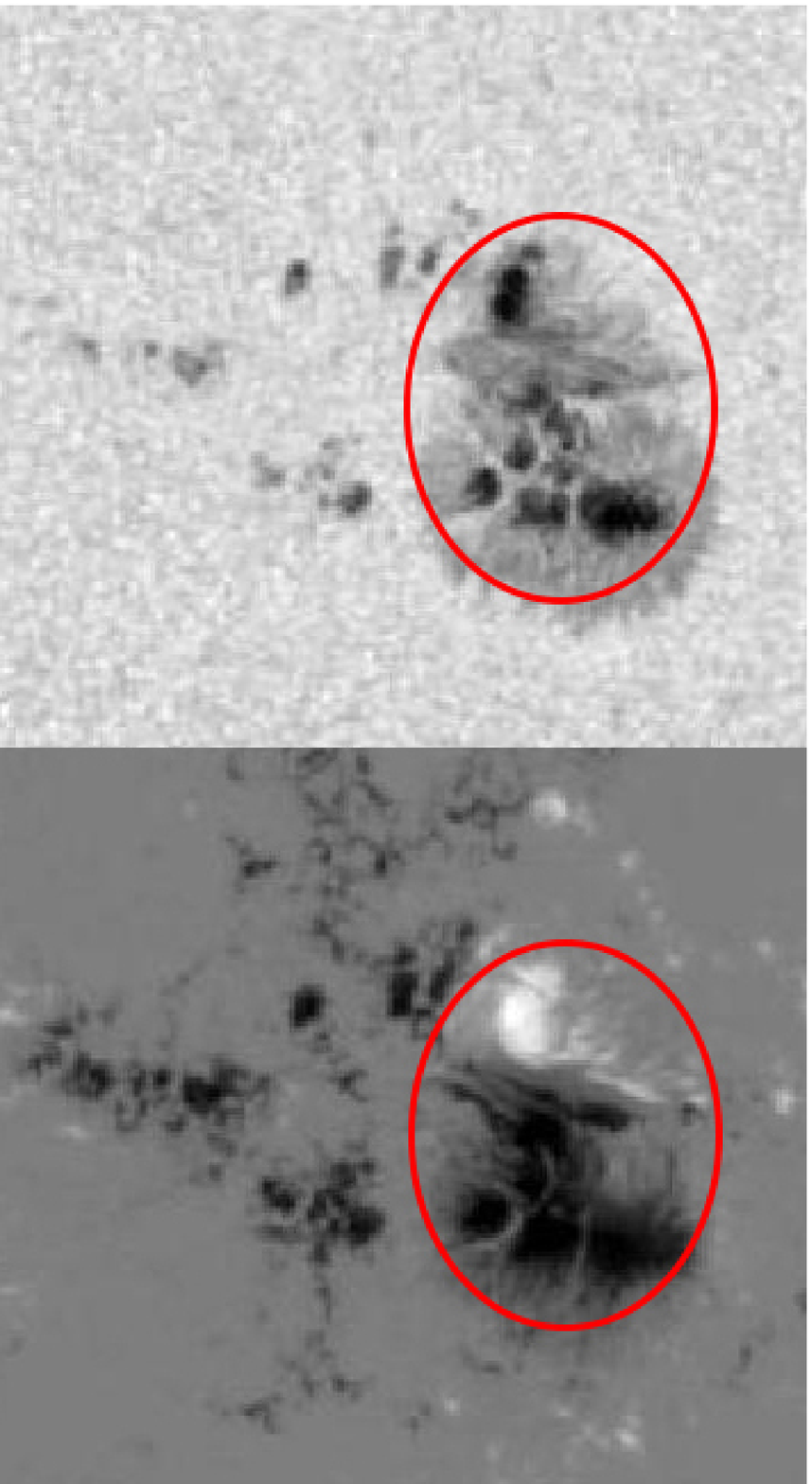}
\put(-110,215){(b)}
\put(-45,180){{\it Area}}
\put(-45,80){{\it Area}}
\put(-45,80){{\it Area}}

\caption{\label{fig11283} Same as Fig. \ref{fig11261} but for of AR 11283.  }
\end{figure*}

\begin{figure*}
\centering
\includegraphics[scale=0.8]{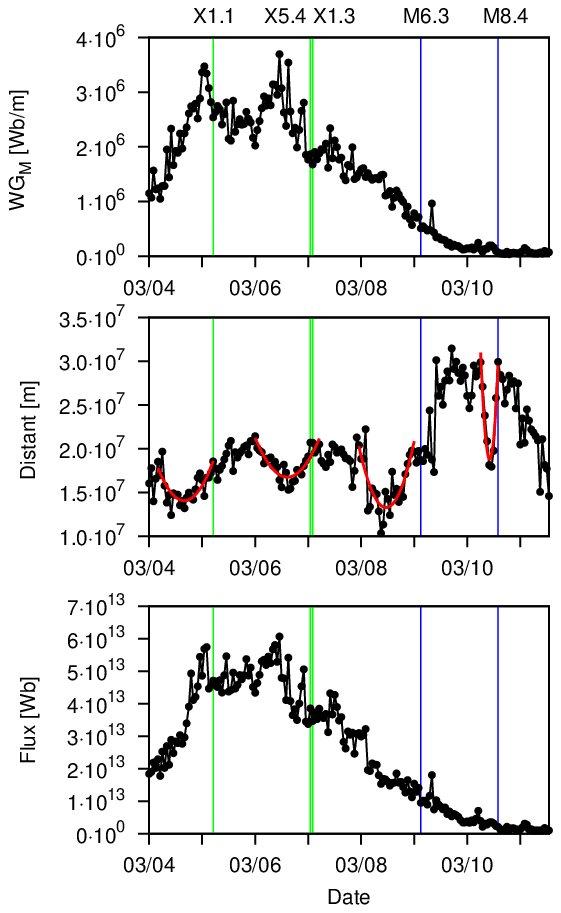}
\put(-135,215){(a)}
\includegraphics[scale=0.25]{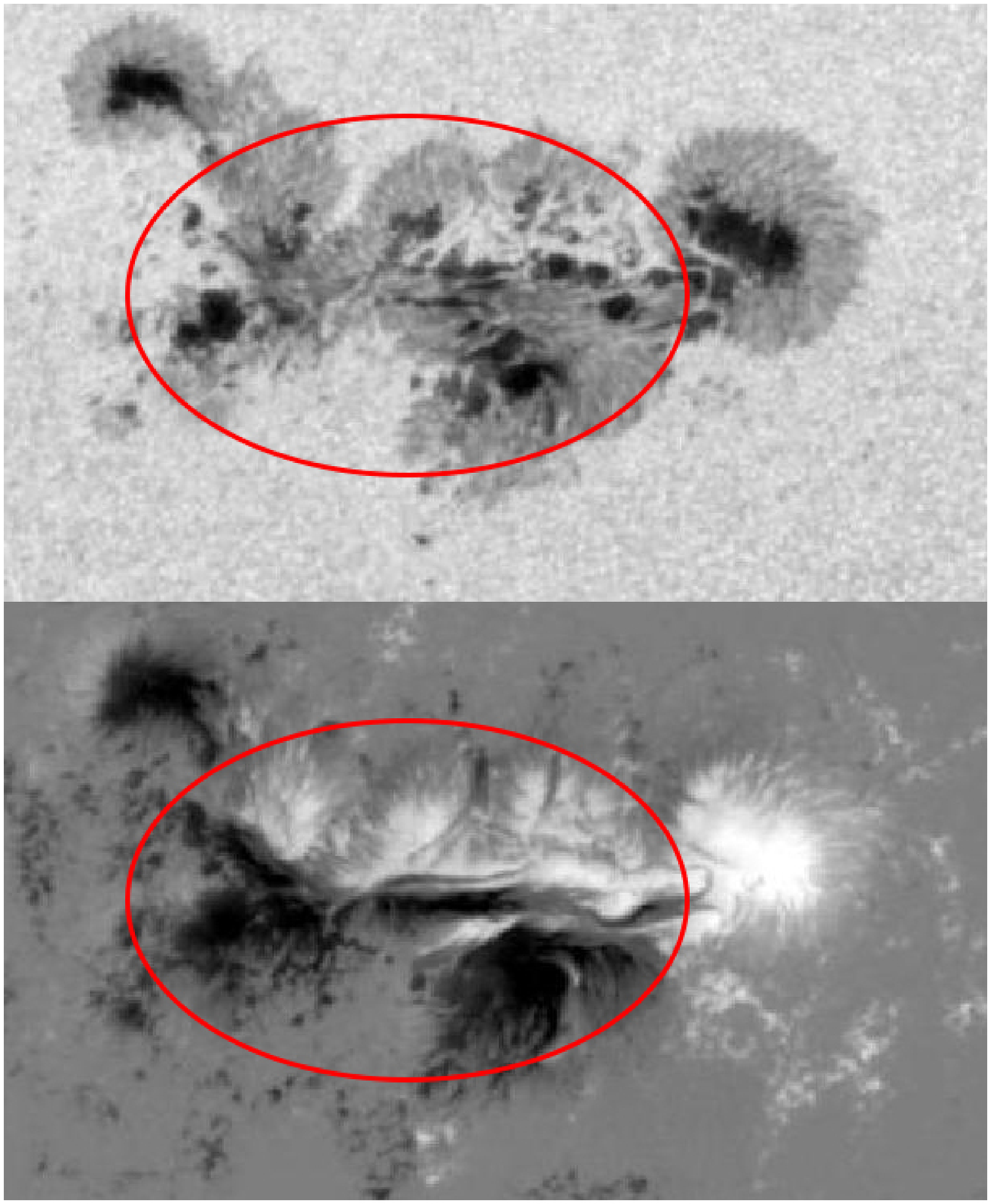}
\put(-155,215){(b)}
\put(-95,190){{\it Area}}
\put(-95,95){{\it Area}}
\put(-95,95){{\it Area}}
\caption{\label{fig11429} Same as Fig. \ref{fig11261} but for AR 11429.  }
\end{figure*}

In the last columns of Figures~\ref{fig11261}, ~\ref{fig11283} and ~\ref{fig11429}, ARs 11261, 11283 and 11429  are shown, in their white-light appearance (upper panel) and the corresponding magnetogram (bottom panel). The red circles highlight the study area for the use of the $WG_{M}$ method. The remaining panels of Figures~\ref{fig11261}-~\ref{fig11429} i.e  the results derived from the  analysis of  the $W G_{M}$, shows the variation of the $W G_M$ (top panel), distance (middle panel), and net flux (bottom panel) over the analysed time series are plotted. In Figures~\ref{fig11261}-~\ref{fig11429}, the vertical blue/green lines mark the M/X-class flares. The column (b) of Fig.~\ref{fig11261} is associated with "Area 1" and column (c) with "Area 2". 

We may state, in general, that the pre-flare behaviour of the weighted horizontal magnetic gradient applied to the three studied ARs, (AR 11261, 11283 and 11429) confirms well and is agreement with the results presented by \cite{Korsos2015}.
Indeed, we can recognised the distinguishing pre-flare behaviour of $WG_{M}$ i.e., that it has a steep rise and a high maximum value followed by a less steep decrease before the flare(s) occurred (upper panels of Figures~\ref{fig11261}-~\ref{fig11429}). Furthermore, by inspecting the middle panels of Figures~\ref{fig11261}-~\ref{fig11429} we observe that the distance parameter shows the unique converging-diverging motion, often referred to as the U-shaped phase, prior to the flare(s) which is a necessary condition for the reconnection processes associated with flares \citep[see][]{Korsos2015}. 

Let us now estimate the predicted maximum flare intensity ($S_{flare}$ in the 1-8 \AA \, wavelength range of GOES) from the maximum value of $WG_{M}$ according to Equation~(1) of \cite{Korsos2016}. The obtained estimated flare classes are in the last but one column of Tables \ref{table1}-\ref{table3} for each AR, respectively. The agreement with the measured GOES classification is acceptable but not best. In most cases only the estimated GOES-class agrees with its measured counterpart. Therefore this tool may require further refinement for a better match. However, this is anyway not the subject of the current work. Next, also estimate the predicted flare onset times ($T_{pred}$) from the variation of the moment of start time of the converging phase ($T_{C}$)  of the distance by the next Equation in all investigated solar eruptions, namely:

\begin{equation}
T_{pred} = a_{1} \cdot T_{C} + b_{1},  
 \label{time}
\end{equation}

where $a_{1} =1.29 (0.85)$ [hr] and  $b_{1} =1.11 (12.8)$ [hr] in the younger (older) than three days case, respectively. In this study, the first M9.3 class-flare of AR 11261 and the M5.3 class-flare of AR 11283 happened before the threshold of 72 hours have elapsed, while, the further 11 investigated flares all occurred after the 72-hr threshold. Therefore for the estimate of the flare onset time of the M9.3 (of AR 11261) and M5.3 (of AR 11283) flares we use $a =1.29$ [hr] and $b =1.11$ [hr]  and for the 11 remain investigated flares, $a =0.85$ [hrs] and $b =12.8$ [hrs] in Equation (\ref{time}).  In general, one can conclude that the Eq. (\ref{time}) over-estimates the flare onset time. 

Tables~\ref{table1}-\ref{table3} summarise the results of applying the $WG_{M}$ method, i.e. we list the various properties of the investigated flares and the accompanied linear velocity ($v_{lin}$) of the CME of the three active regions (AR11261, AR11283 and AR11429). Furthermore, Tables ~\ref{table1}-\ref{table3} also include the maximum value of the $WG_{M}$, value of $WG_{M}$ at the flare onset, the duration of the observed compressing phase ($T_{C}$), elapsed time between the minimum point until flare onset ($T_{D+F}$), the predicted flare onset time ($T_{pred}$ computed by Equation (~\ref{time})), the predicted flare intensity \citep[$S_{flare}$ determined by Equation (1) of ][]{Korsos2016} and the ratio of maximum value of the $WG_{M}$ and the value of the $WG_{M}$ at the flare onset. The ratio is also an important diagnostic tool of the $WG_{M}$ method, because as discussed in \cite{Korsos2015}, we found that if the value of $WG_{M}$ decreases over than $\sim$54\% after the local maxima of $WG_{M}$ then no further energetic flare(s) can be expected; but, if the maximum of the released flare energy is less than about 42\%, further flares are more probable.

In briefly, we can conclude that the $WG_{M}$ method is fairly estimated the expected flare intensity and the estimated onset time. However, the flare-prediction capability of the $W G_M$ method could be further 
improved, by analysing other physical quantities of flaring ARs. Therefore we embark on investigating the evolution of the total, shearing and the emerging helicity of ARs before the flare and CME eruptions for these three cases studies.

\begin{table*}
\caption { The examined properties of the AR 11261}
\resizebox{\textwidth}{!}{
\begin{tabular}{ccccccccccc}
\hline

Flare &Flare onset time& $v_{lin}$ of CME&Maximum $WG_{M}$  &Onset $WG_{M}$  & $T_{C}$  &$T_{D+F}$   &   $T_{pred}$ & $S_{flare}$&Decrease\\
&& [km/s]&$\cdot$$10^{6}$ [Wb/m]  &$\cdot$$10^{6}$ [Wb/m] & [hour] & [hour] &   [hour] & & [\%]\\

\hline
M9.3   &30/07/2011 02:02& - &    2.7	&  1.2		& 11	&  12	 & 21.43	&M9.9&55\%	 \\
M1.4    &02/08/2011 06:24& 712 &   2.5 	&  2.0	&11 & 6  &21.43 &	M9.1&22\%		 \\	
M6.0    & 03/08/2011 13:54&610&  2.5	&  	1.9	&9 & 13  &19.73 &	M9.1&27\%		 \\	
M9.3     &04/08/2011 04:09&1315 &   1.5	&  1	&29& 17  &36.73	& M5.4&36\%		 \\

\hline

\hline
\end{tabular}}

\label{table1}
\end{table*}

\begin{table*}
\caption { The examined properties of the AR 11283}
\resizebox{\textwidth}{!}{

\begin{tabular}{ccccccccccc}
\hline

Flare &Flare onset time& CME&Maximum $WG_{M}$  &Onset $WG_{M}$  & $T_{C}$  &$T_{D+F}$   &   $T_{pred}$ & $ S_{flare}$&Decrease\\
&&$v_{lin.}$ [km/s]&$\cdot$$10^{6}$ [Wb/m]  &$\cdot$$10^{6}$ [Wb/m] & [hour] & [hour] &   [hour] & & [\%]\\

\hline
M5.3   &06/09/2011 01:35& 782&  0.7  	&  1.0		& 13	&  30 	 & 	17.88	&M3.0&-	 \\
X2.1    &06/09/2011 22:12&575 &  1.1	&  0.8		&17 & 	14 &26.53  &M3.8	&24\%		 \\	
X1.8    & 07/09/2011 23:10&792&   1.4	&  1.0		&17 & 	39  &26.53 &M5.2	&26\%		 \\	
M2.7     &09/09/2011 07:10&318&  1.4  	&  0.8		&17 & 	70  &26.53& M5.2	&43\%		 \\

\hline

\hline
\end{tabular}}

\label{table2}
\end{table*}

\begin{table*}
\caption { The examined properties of the AR 11429}
\resizebox{\textwidth}{!}{

\begin{tabular}{ccccccccccc}
\hline

Flare &Flare onset time& CME&Maximum $WG_{M}$  &Onset $WG_{M}$  & $T_{C}$  &$T_{D+F}$   &   $T_{pred}$ & $S_{flare}$&Decrease\\
&&$v_{lin.}$ [km/s]&$\cdot$$10^{6}$ [Wb/m]  &$\cdot$$10^{6}$ [Wb/m] & [hour] & [hour] &   [hour] && [\%]\\

\hline
X1.1   &05/03/2012  04:30& 1531&   3.5 	&2.5  & 10	&  11 	 & 	20.58	&X1.2&27\%	 \\
X5.4    &07/03/2012  00:02&2684 &    3.7	&  1.9&12 & 7  &22.28&	X1.3&48\%		 \\	
X1.6    & 07/03/2012  01:14&1825&3.7  	&  1.8&12 &8 &23.98 	&X1.3&52\%		 \\	
M6.3     &09/03/2012  03:22&950& 3.7  	&  0.5&11 & 13 &21.43 	&X1.3&86\%		 \\	
M8.4     &10/03/2012  17:15&1296&  3.7  	&  0.01	&5 & 4  &16.33 &X1.3	&98\%		 \\

\hline

\hline
\end{tabular}}

\label{table3}
\end{table*}

 \subsection{Magnetic Helicity Method and Application to 3 different ARs case}

Let us now determine the magnetic helicity associated with the three ARs each, and, investigate their evolution prior the eruptions. The magnetic helicity flux across a surface $S$ introduced by \cite{Berger1984} can be expressed as:

\begin{equation} \label{Origin}
\left. \frac{dH}{dt}\right|_{S}=2\int_{S}\left( {\bf A}_{p}\cdot {\bf B}_{h}\right){\bf v}_{\bot z}dS-2\int_{S}\left( {\bf A}_{p}\cdot {\bf v}_{\bot h}\right) {\bf B}_{z}dS,
\end{equation}

 where ${\bf A}_{p}$ is the vector potential of the potential field. ${\bf B_{p}}$, ${\bf B}_{h}$ and ${\bf B}_{z}$ denote the tangential and normal magnetic fields, and ${\bf v}_{\bot h}$ and ${\bf v}_{\bot z}$ are the tangential and normal components of velocity ${\bf v}_{\bot}$. The first term on the right side of Equation (\ref{Origin}) is the helicity generated from shearing motions while the second term is the helicity from emerging motions. ${\bf A}_{p}$ is determined by the photospheric vertical magnetic field and Coulomb gauge by equations  \citep{Berger1997,Berger2000}:

\begin{equation}
\nabla \times {\bf A}_{p} \cdot {\bf{\hat{n}}} = B_{h}, \nabla \cdot {\bf A}_{p} = 0, {\bf A}_{p} \cdot {\bf{\hat{n}}}=0.
\end{equation}

Based on the basic algebraic relations, we then obtain:

\begin{eqnarray} \label{velocity}
v_{\parallel}=\dfrac {\left( {\bf v}\cdot {\bf B}\right){\bf B}}{B^{2}},              \\
{\bf v}_\bot={\bf v}-\dfrac {\left( {\bf v}\cdot {\bf B}\right){\bf B}}{B^{2}},         \\
{\bf v}_{\bot h}={\bf v}_h-\dfrac {\left( {\bf v}\cdot {\bf B}\right){\bf B}_h}{B^{2}}, \\
{\bf v}_{\bot z}={\bf v}_z-\dfrac {\left( {\bf v}\cdot {\bf B}\right){\bf B}_z}{B^{2}}.
\end{eqnarray}

Here, ${\bf v}$ is the photospheric plasma velocity, ${\bf v}_{\parallel}$ and ${\bf v}_{\bot}$ denotes the velocity components that is parallel and perpendicular to the magnetic field.

 Solar Dynamic Observatory (SDO) was launched in 2010. The on-board Helioseismic and Magnetic Imager (HMI) can map the full disk photospheric vector magnetic field with high cadence and long continuity. The vector magnetograms employed in this study are from Space-weather HMI Active Region Patches (SHARPs) with a spatial resolution of 1" and a 12 mins cadence \citep{Bobra2014}. The photospheric plasma velocity was calculated using the Differential Affine Velocity Estimator for Vector Magnetograms (DAVE4VM) algorithm \citep{Schuck2008}, the window size used in the calculation is 19 pixels, which was determined by examining non-parametric Spearman rank order correlation coefficients, Pearson correlation coefficients and slopes between $\Delta_{h} \cdot ({\bf v}_z{\bf B}_h-{\bf v}_h{\bf B}_z)$ and $\delta {\bf B}_z/\delta t$ \citep{Schuck2008}. The vector potential ${\bf A}_{p}$ is derived using MUDPACK \citep{Adams1993}, a multigrid software for solving elliptic partial differential equations. Then, we calculated magnetic helicity from these active regions using Equation (\ref{Origin}). The helicity injection rate could be obtained by integrating over the entire calculation area. Magnetic helicity generated by shearing motion and emerging motion are calculated separately, and the total helicity is these combination of the two components.

 \begin{figure*}
 \centering
 \includegraphics[width=0.47\linewidth]{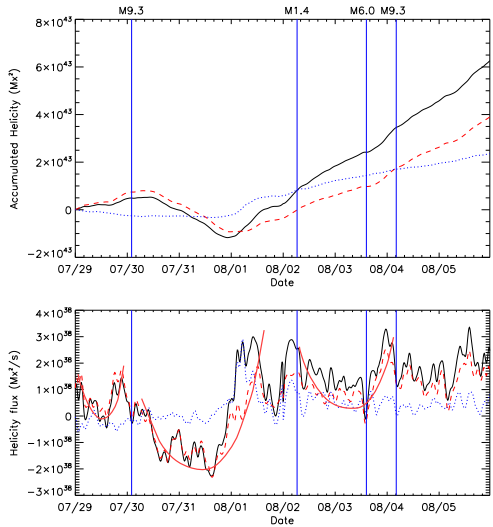}
 \caption{\label{Heicity11261}  Top panel: Accumulated helicity from AR 11261; Bottom panel: Helicity flux of AR 11261. The red dashed line is the helicity from shearing motion, the blue dotted line is the helicity form emerging motion, and the black solid line is the total helicity. The red parabolae highlight the decreasing-increasing phases similar to a feature found in the $WG_{M}$ results.}
 \end{figure*}

 \begin{figure*}
 \centering
 \includegraphics[width=0.47\linewidth]{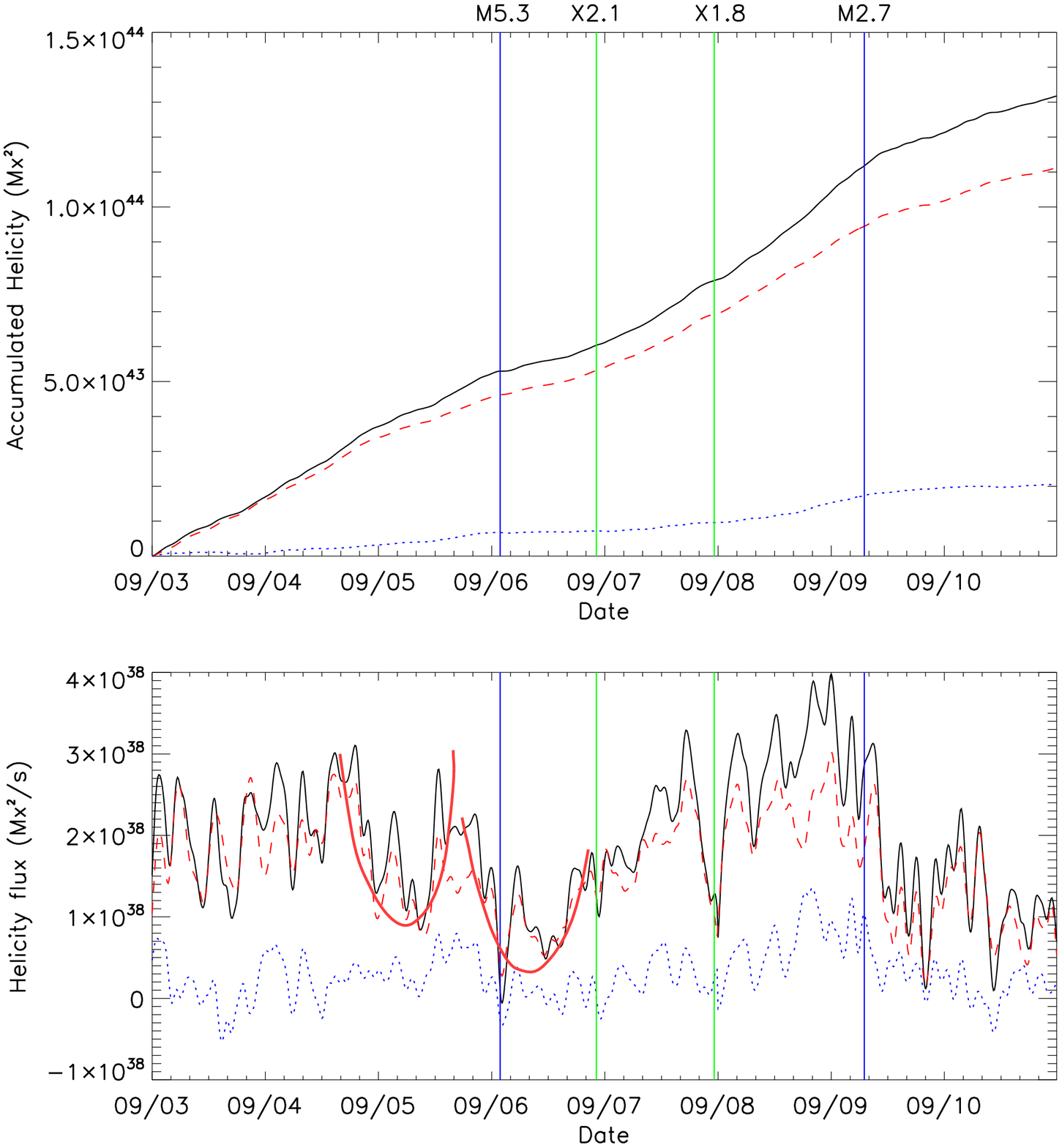}
 \caption{\label{Heicity11283}  Same as Fig. \ref{Heicity11261} but for AR 11283.}
 \end{figure*}

 \begin{figure*}
 \centering
 \includegraphics[width=0.47\linewidth]{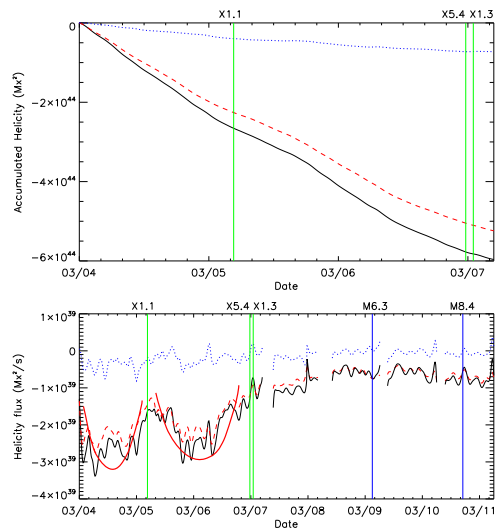}
 \caption{\label{Heicity11429} Same as Fig. \ref{Heicity11261} but for AR 11429. }
 \end{figure*}

Temporal profiles of helicity fluxes in the three ARs are plotted in Figures \ref{Heicity11261}, \ref{Heicity11283} and \ref{Heicity11429}. In each figure, the bottom panel is the helicity flux, while the red dashed line is the magnetic helicity flux generated by shearing and twisting movements at the photosphere, the blue dot line stands for that transported across the photosphere, and the black solid line is the total magnetic flux. The top panel shows the accumulated helicity which is obtained by integrating the helicity flux from the start of the observation to the specified time.

As there are several big gaps in SHARP's data from March 3 to March 11 while with no vector magnetograms, helicity fluxes in Figure \ref{Heicity11429} had been separated into five segments and the accumulated helicity's time sequence was calculated from 00:00:00 UTC, March 4 to 06:36:00 UTC, March 7.

The magnetic helicity flux shows a decrease before every M-class or above flares  \citep[see, e.g.,][and references therein]{Smyrli2010}. In AR 11261, the magnetic helicity decrease from the pre-flare highest time to the flare on-set time is $1.7\times 10^{38} \ \mathrm{Mx^{2}s^{-1}}$, which is $3.6\times 10^{37} \ \mathrm{Mx^{2}s^{-1}}$, $1.6\times 10^{38} \ \mathrm{Mx^{2}s^{-1}}$ and $2.1\times 10^{38} \ \mathrm{Mx^{2}s^{-1}}$ for the following ARs. In AR 11283, the decrease before each AR is $1.7\times 10^{38} \ \mathrm{Mx^{2}s^{-1}}$, $8\times 10^{37} \ \mathrm{Mx^{2}s^{-1}}$, $2.1\times 10^{38}  \ \mathrm{Mx^{2}s^{-1}}$ and $1.3\times 10^{38}  \ \mathrm{Mx^{2}s^{-1}}$. The total helicity flux in AR 11429 is negative, the absolute value has a decrease of $1.2\times 10^{39}  \ \mathrm{Mx^{2}s^{-1}}$ in the first X1.1 flare, and a total of  $1.8\times 10^{39} \ \mathrm{Mx^{2}s^{-1}}$ change in the following two X-class flares. With three X-class flares and two M-class ones being produced in AR 11429 and a corresponding much higher helicity injection than that in the other two ARs (AR 11261 and 11283), it suggests that large helicity flux which injects magnetic free energy continuously into the solar atmosphere may results in fierce flare eruptions. Also, the magnetic helicity flux from emerging motion is more stable than that from shearing motion which fluctuates considerable during the AR's life time. It also can be found that before large flares, the helicity flux from shearing motion dominated the helicity accumulation, which indicates its essential position in solar eruptions.

 Besides, several clear long-duration decreasing-increasing phases could be found in either the total helicity flux or the shearing helicity flux before large flares, even covered a day during the entire phase. Strong shearing movement along the PIL introduced large shearing helicity fluxes with opposite sign in both sides of the PIL resulting in a downward trend in the total shearing helicity flux in the entire area of interest. When such shearing motion became weakened, the total shearing helicity would increase with the domination of one polarity's helicity flux.

\section{Conclusion } \label{conclusion}

There are several flare-forecasting methods based on photospheric observations of the ARs in the solar atmosphere. Here, we applied two different approaches to develop the basics of a good and reliable flare and CME prediction model \citep[for a review see][and references therein]{Benz2017, Leka2017}. One approach is the weighted horizontal magnetic gradient method (denoted as $WG_{M}$) introduced in \cite{Korsos2015} and the second one is employing magnetic helicity \citep{Berger1984,Berger1984b}. We tested these two methods in three different flare- and CME-rich ARs, namely AR 11261, 11283 and 11429. All three active regions produced a series of solar eruptive occurrances. AR 11261 hosted four M-class flares where one was accompanied with a fast CME. AR 11283 had similar activity than AR 11261 with two M- and two X-class flares, however, only with a slow CME. Finally, AR 11429 was the most powerful of the three active regions as it gave birth to five very compact solar eruptions.

Applying the first investigation, we follow the temporal evolution of $WG_{M}$ and the distance between the area-weighted barycenters of opposite polarities within an appropriately defined region close to the magnetic polarity inversion line (PIL)  of the three studied ARs. During the empirical analyses of the three ARs, first, we recognised typical pre-flare behaviour patterns of $WG_{M}$ and distance likewise in \cite{Korsos2015}. One remarkable behaviour of opposite polarities is that, there is indeed the steep rise, and the maximum value of the magnetic flux gradient is followed by a less steep decrease before the flare and CME occurrences.  Parallel to the increasing/decreasing trends of $WG_{M}$, concurrent decreasing/increasing (approaching/receding) trends of distances, called as U-shaped pattern, were also observed for opposite polarity spots. 

The second approach is employing the total magnetic helicity calculation. We separately followed the evolution of the total, emerging and shearing helicity components prior the flare and CME occurrences. In general, the total magnetic helicity is divided into two terms, one is from the emergence of twisted field lines that cross the photospheric surface (this is the so-called emerging helicity) and the other one is from the shearing motion in the photosphere that are twisting field lines (this is where the shearing helicity comes from) \citep{Berger1984,Berger1984b}.

In the helicity calculation, we recognized similar decreasing -increasing phases in the evolution of shearing and total helicity before the flare(s) and CME(s) occurred, just as found for the decreasing-receding phase of flares when applying $WG_{M}$. This common property is highlighted by red parabolae in Figures~\ref{fig11261}--~\ref{Heicity11429}.  We can also conclude that the duration of decreasing-increasing phases are very comparable during the evolution of shearing helicity and distance between the area-weighted barycenters of the spots of opposite polarities. Furthermore, we note that we cannot determine any meaningful behaviour in the evolution of the emerging helicity.  Therefore, it is worth pointing out that the shearing motions may play a more  important role in the formation of total helicity because the value of emerging helicity is negligible when compared to the value of the shearing helicity. 

According to our empirical case studies, we can clearly identify a {\it common decreasing-increasing phase} in the evolution of {\it shearing helicity} and {\it weighted distance} prior to flare and CME eruptions (see Figs.~\ref{fig11261}--~\ref{Heicity11429}). This new result is really interesting, because we may conclude that the shearing mechanism may also be an equally key driver for the solar eruption, and perhaps not only the so much studied emerging process is relevant.
In the literature, there are several flare and CME models based on the photospheric shear motion. \cite{Sturrock1966} introduced the tearing-mode instability model which is based on the shearing motion at the photosphere, or there is the model of sheared loops inside arcade by \cite{Somov1998}. The magnetic breakout model, presented by \cite{Antiochos1999} is also based on photospheric shearing motions. But, in the literature, the emerging flux process seems to be more favoured and more acceptable \citep[see, e.g.,][and references therein]{Chandra2009,Takafumi2014,Louis2015} when trying to understand flare/CME eruption dynamics.

We would also emphasise that we do not say that the emerging process is not needed for analysing or predicting large-scale solar eruptions. On the contrary, without flux emergence there is, of course, likely no flaring. Finally, we argue that there is a need for a much larger statistical study in order to confirm our conjecture formulated in this work. However, this is beyond the scope of the present case studies.

    \section{ Acknowledgements}
       We acknowledge the use of HMI SHARPs data from SDO/HMI team. This work is supported by the National Science Foundation of China (NSFC) under grant number 41231068. YDY is grateful for the support from the State Key Laboratory of Space Weather, National Space Science Center, CAS. The author MBK is grateful to the University of Sheffield for the support received while carrying out part of the research there.  RE is grateful to Science and Technology Facilities Council (STFC) UK and the Royal Society (UK). The authors also acknowledge the support received from the CAS Key Laboratory of Solar Activity, National Astronomical Observatories  Commission for Collaborating Research Program. RE acknowledges the support received from the CAS Presidents International Fellowship Initiative, Grant No. 2016VMA045.

\end{document}